\documentclass[namedreferences]{solarphysics}

\usepackage[hyperref,optionalrh]{spr-sola-addons} 
\usepackage{graphicx}        
\usepackage{color}           
\usepackage{breakurl}        

\begin{document}

\begin{article}
\begin{opening}

\title{Reconstruction of the Solar Activity from the Catalogs of the Zurich Observatory}

\author[addressref={aff1,aff2},email={egor.illarionov@math.msu.ru}]{\inits{E.A.}\fnm{Egor}~\lnm{Illarionov}\orcid{0000-0002-2858-9625}}
\author[addressref=aff3,email={}]{\inits{R.}\fnm{Rainer}~\lnm{Arlt}
}
\address[id=aff1]{Moscow State University, Moscow, Russia}
\address[id=aff2]{Institute of Continuous Media Mechanics, Perm, Russia}
\address[id=aff3]{Leibniz Institute for Astrophysics, Potsdam, Germany}

\runningauthor{Illarionov \& Arlt}
\runningtitle{\textit{Solar Physics} Reconstruction of the solar activity}

\begin{abstract}
Catalogs of the Zurich Observatory contain positional information on sunspots, prominences and faculae in late 19th and early 20th centuries. This database is given in handwritten tabular form and was not systematically analysed earlier. It is different from the sunspot number time series made in Zurich and was obtained with a larger telescope. We trained a neural-network model for handwritten text recognition and present the database of reconstructed coordinates. The database obtained connects the earlier observations by Sp\"orer with later programs of the 20th century and supplements the sunspot-group catalogs of the Royal Greenwich Observatory. We also expect that the presented machine-learning approach and its deep capabilities will motivate the processing of a wide bulk of astronomical data, which is still given in non-digitized form or as plain scanned images.
\end{abstract}

\end{opening}

\section{Introduction}
Telescopic observations of sunspots are available for more than 400 years. 
The increasing amount of digitized versions of historical books and
manuscripts has propelled the detailed analysis of those observations
in the last decade. An overview of the utilization of historic sunspot drawings 
in particular was given by \citet{arlt_vaquero2020}.

A long time and frequently used series of sunspot data is available from
photographs of the Royal Greenwich Observatory for 1874--1976. The data
are actually a compilation of results from several observatories around 
the world as to cover nearly every single day of the 103-year period
\citep{willis_ea2013}. The disadvantage of the catalogue is that no
information on individual spots is available, but only average positions
of sunspot groups as well as their areas. Moreover, it has been argued, e.g. by \citet{cliver_ea2014} and \citet{cliver2017}, that the Greenwich catalogue has a deficit of groups until about 1915 as compared to the rest of the time series.

The period around 1900 saw a relatively low Cycle~14 and may be associated
with a secular minimum which solar activity encounters roughly once a century.
A detailed knowledge of the various manifestations of solar activity in
those periods when activity is lower for many years is interesting as to what 
is different compared to other cycles.

The present paper attempts to bring the sunspot positions of the Zurich Observatory
into digital form. The full period covered with a consistent analysis method
is from 1883--1936. The observations are based on direct measurements of
coordinates and are accompanied by detailed drawings of the Sun.

Previous attempts of pattern recognition on graphical sunspot data include, e.g., 
the reconstruction of the Carrington catalog by \citet{Lepshokov2012}, the analysis
by \citet{diercke_ea2015} of sunspot positions in the synoptic maps by Gustav Sp\"orer,  
and the partial reconstruction of the synoptic maps made by Alfred Wolfer in Zurich
\citep{Tlatova}.  On the more contemporary side, MDI images were analysed through 
automated recognition by \citet{watson_fletcher2011}. Data utilization by character 
recognition on printed tables was performed by \citet{curto_ea2016}, for instance.

\section{Data Set\label{dataset}}
When Alfred Wolfer started working under Rudolf Wolf at Eidgen\"ossische
Sternwarte Z\"urich in Switzerland, he started a programme with detailed
drawings of all sunspots, faculae and prominences for as many days as
possible. The time series of Wolf numbers was continued as a separate
programme with a smaller telescope, in order to keep the consistency of
the numbers with the past. In the positional programme, most spots, faculae,
and prominence were measured with their locations and compiled in result
books, available as digital documents at e-manuscripta in 
Zurich\footnote{\url{https://www.e-manuscripta.ch/}}. We list the permanent links
to the individual books in Table~\ref{table_links}. Wolfer reported that
he actually started with positional measurements in the autumn of 1879
and gives sunspot positions from August~1879 to June~1880 in \citet{wolf1881}.

Our goal is to convert the images of the scanned, hand-written result books 
into numerical tables with the positions of solar activity
features. The book for 1885--1886 is missing in the digital library of
e-manuscripta. There are also no individual drawings from those years in the 
digital archive. Alfred Wolfer gives an instrumental description 
as well as an explanation for the break in 1885--1886 in \citet{wolf1887}, 
which we translate to 
\begin{quote}``After the positional determinations of sunspots had
been interrupted in the end of 1884 because of the planned modification of the
refractor, I have now again started the observations this January with the
resurrection of the latter, and the observations are still the main task for
this instrument.''
\end{quote}
The first book of 1883--1884 contains only spot information, while the second
book of 1887--1892 begins with positions of spots and faculae. On July~17,
1888, also prominences were recorded. Since then, sunspot positions are 
given in black ink, positions of faculae are given in blue ink, and positions 
of prominences are in red. We will come back to the colors in 
Section~\ref{color_identification}.

There is a brief interruption of the observations in 1891, from July~31 to
September~19, because 54~pages are missing in the digital version of the
corresponding observing book.

Individual sunspots were measured, except when forming very small clusters
of two or more spots. Such clusters are much smaller than the typical size
of a sunspot group. Also, two umbrae in a common penumbra are usually
given as one spot. Typical abbreviations in the tables of positions are
`kl' for `klein'~=~small, `Gr kl Fl' for `Gruppe kleiner Flecken'~=~
group of small spots, `Hof\/fl' for `Hof\/fleck'~=~spot with penumbra, 
`kl Hof\/fl' for `kleiner Hof\/fleck'~=~small spot with penumbra, `Kernfl'
for `Kernfleck'~=~penumbra with several umbrae. Sometimes, remarks like
`ohne Hof'~=~without penumbra appear.

\begin{table}
\caption{References to the twelve books of positions of solar activity features, available at e-manuscripta. Books listed in italics are not part of the present paper.\label{table_links}}
\begin{tabular}{ll}
\hline
Years & Reference \\
\hline
{\em1883--1884} & {\em https://doi.org/10.7891/e-manuscripta-48582} \\
1887--1892 & https://doi.org/10.7891/e-manuscripta-48586 \\
1893--1898 & https://doi.org/10.7891/e-manuscripta-48585 \\
1899--1902 & https://doi.org/10.7891/e-manuscripta-48587 \\
1903--1905 & https://doi.org/10.7891/e-manuscripta-48583 \\
1906--1910 & https://doi.org/10.7891/e-manuscripta-48850 \\
1911--1915 & https://doi.org/10.7891/e-manuscripta-48849 \\
1916--1920 & https://doi.org/10.7891/e-manuscripta-48848 \\
{\em 1921--1925} & {\em https://doi.org/10.7891/e-manuscripta-48851} \\
{\em 1926--1930} & {\em https://doi.org/10.7891/e-manuscripta-49040} \\
{\em 1931--1935} & {\em https://doi.org/10.7891/e-manuscripta-49038} \\
{\em 1937--1938} & {\em https://doi.org/10.7891/e-manuscripta-49039} \\
\hline
\end{tabular}
\end{table}

In \citet{wolf1881}, which is number~53 of his famous series of sunspot
reports (`Mittheilungen'), Alfred Wolfer describes the measurement
technique. The telescope was a 16-cm refractor with a focal length
of 2.63~m. The solar image was projected onto a white screen behind 
the eyepiece and delivered a diameter of about 50~cm. Cross-hairs in
the eyepiece, which are projected as well, are used to determine 
right ascension and declinations distances from the western, eastern,
northern and southern solar limbs, while the right ascension differences
are time measurements and the declination differences are obtained with
a micrometer screw.

The conversion of coordinates follows \citet{spoerer1874} and goes through polar 
coordinates on the solar disk to heliocentric ecliptical coordinates, which are then
transformed to heliographic coordinates using an ephemeris of the solar axis.
Since this was computationally expensive at the time, auxiliary tables were
used for various steps in which the observers and assistants interpolated
`by eye'. Whereas the atmospheric refraction is automatically taken into
account by using measurements from top and bottom, differential refraction
is not, i.e. the change of refraction across the solar disk, and the physical
deviation of the Sun from a perfect sphere is, of course, also not taken
into account, but it is extremely small (less then 10 milli-arcseconds).

\begin{figure}
\centering
\includegraphics[width=\textwidth]{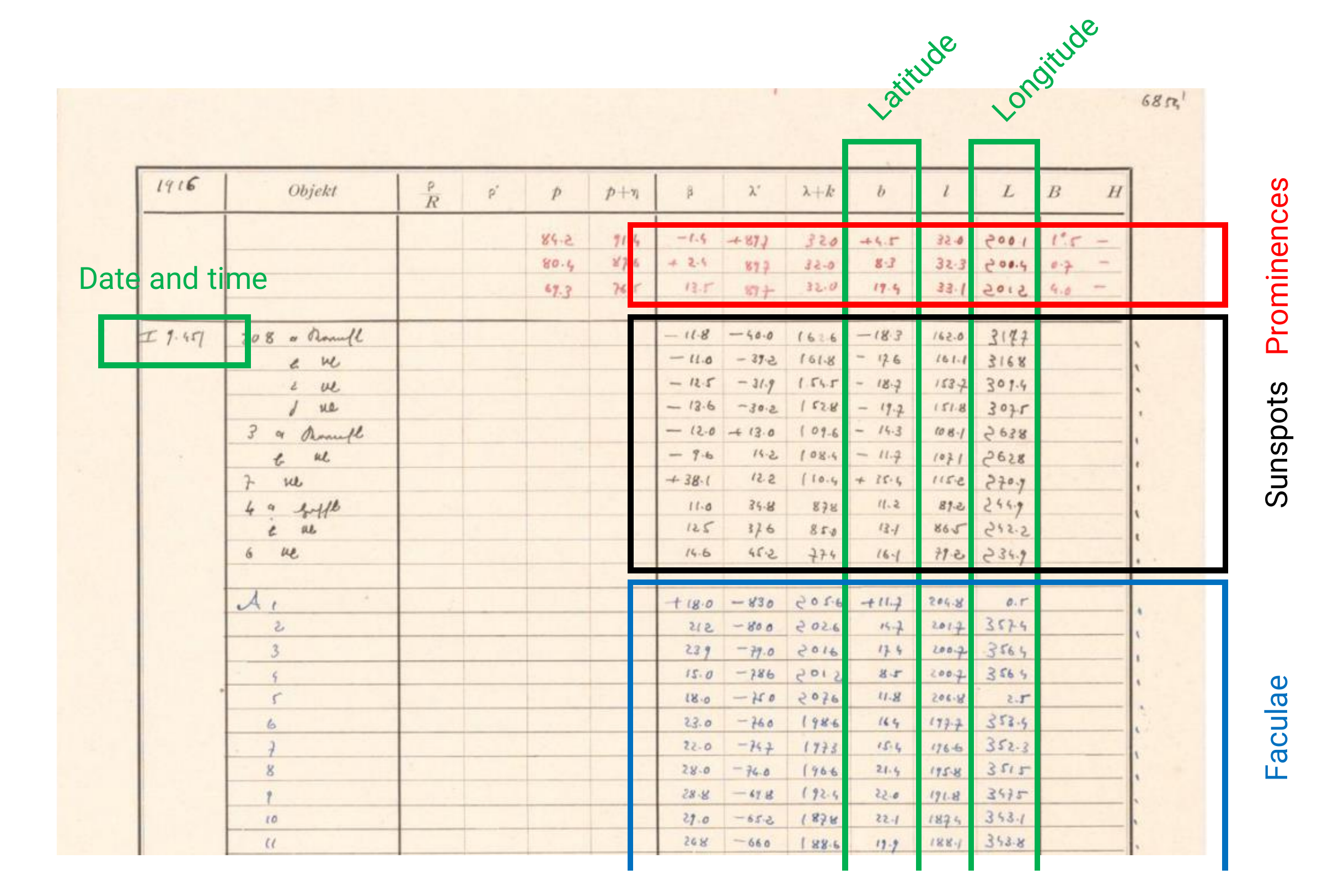}
\caption{A sample page showing the key elements of the page layout (only the top half of the page is shown). Green boxes show date and time of the observation and the two columns with heliographic latitude and longitude. The red box contains coordinates of prominences, the black box is for sunspots, the blue box is for faculae. }
\label{fig:layout}
\end{figure}

Figure~\ref{fig:layout} shows a sample page layout. The key elements of the page are date and time of the observations and 6 columns with coordinates labeled as $\beta$, $\lambda'$, $\lambda+k$, $b$, $l$ and $L$. We interpret $b$ and $L$ as heliographic latitude and longitude. The remaining columns represent coordinates in auxiliary coordinate systems. In the beginning of the observing programme, also the polar coordinates were reported; they are omitted in most of the books though.

The heliographic longitude $L$ is a sidereal value obtained at the
time of writing the observing books, using a fixed angular velocity
of $14.2665$~degrees/day \citep{wolf1881} which corresponds to a sidereal rotation
period of $25.2339$~days. Note that this value is slightly smaller
than the value used by, e.g., the JPL Horizons ephemeris service\footnote{\url{https://ssd.jpl.nasa.gov/horizons/}} which is
$25.38$~days.

There are certain relationships between the columns. Thus, for each particular date and time the sum of the columns $\lambda'$ and $\lambda+k$ is constant modulo $360$ as well as the difference of the columns $L$ and $l$ is constant modulo $360$. Also $\sin b = \sin\beta\sin(\omega+i)/\sin\omega$, where $\tan\omega=\tan\beta/\cos(\lambda+k)$ and $i\approx6.96^{\circ}$ is the inclination of the solar ecliptic \citep[see][for derivation of these equations]{spoerer1874}. The above relationships between the columns will play a crucial role in fine-tuning of the numbers recognition model. 

It can be noted in Figure~\ref{fig:layout} that all numbers have a single-decimal precision. This is valid for all catalogs except a part of the earliest ones where sunspot coordinates were given with two-decimal precision. Due to this fact we do not include the dot symbol into the recognition model.

\section{Recognition Method}

The general pipeline is as follows:
\begin{itemize}
    \item Detection of page layout (vertical and horizontal lines that form the table);
    \item Initial recognition of the numbers in table cells using the pre-trained neural network model;
    \item Font color identification;
    \item Derivation of linear relationships between the columns $\lambda'$, $\lambda+k$, $l$ and $L$;
    \item Fine-tuning the neural network model to improve the recognition accuracy;
    \item Post-processing the outputs of the the neural network model to derive the most probable interpretation.
\end{itemize}
    
Below we describe each step in more details.

\subsection{Detection of Page Layout}

To recover the page layout, we need to detect all vertical and horizontal lines and distinguish between thick and thin lines that separate series of observations and individual observations. Note that position and intervals between the lines are not uniform thus each page should be processed individually. 

We first convert the page to grayscale, then apply the Canny filter \citep{Canny86} to detect edges in the image. We eliminate fuzzy edges by smoothing the image with a Gaussian filter. For horizontal lines detection, we then convolve the image with a horizontal kernel and summarize the obtained pixel intensities along image rows. In the obtained array, horizontal lines appear as pronounced peaks that can be easily identified. Of course, one has to carefully adjust filter's strength, kernel sizes and peak's threshold to obtain a robust procedure. Fortunately, a single set of parameters works fine for most of the pages.

Once the page layout is defined, we can isolate (crop) individual cells of the table that will be used as inputs into the text recognition model. For convenience, we also rescale all cell's images to a standard size of $64\times32$ pixels. 

\subsection{Initial Model for Text Recognition}

We apply the convolutional recurrent neural network model (CRNN) designed for sequential text recognition \citep{crnn}. The model takes an image (in our case, of size $64\times32$ pixels) and outputs a 2D array of size $64\times12$ that will be referred as a \textit{heatmap}.

More detailed, the neural network model consists of 3 blocks: sequence of convolutional and downsampling layers, then we have two bidirectional recurrent layers and the final linear (fully connected) layer. The convolutional layers transform input tensor of size (1, 32, 64)\footnote{For neural networks, tensor dimensions are (channels, height, width). } to a tensor of size (128, 1, 64). We roughly interpret it as 128 features attributed to each image column. Then we apply the recurrent layers to obtain a tensor of size (512, 64) that is interpreted as a sequence (of size 64) of feature vectors (of size 512) generated from the sequence of image columns (note that one-to-one correspondence between elements of this sequence and image columns is not assumed here). Finally, the output layer transforms 512~features to~12 that represent  class scores, where classes are `0', `1', $\dots$, `9', the minus sign `$-$' and a special \textit{blank} symbol `$\ast$', which is aimed to separate the main symbols. These 12~symbols form the minimal sufficient set for decoding a number. Sequence of size 64 of 12 class scores forms the heatmap of size $64\times12$.

Note that we do not learn the dot symbol and the plus symbol. The reason is that the plus symbol is a rare symbol with respect to digits and the minus symbol (in fact, plus symbols arise when the initial minus symbol was written by mistake and thus was canceled). Rare symbols will cause a high class imbalance during the model training and most likely result in low accuracy for this symbol. The dot symbol is rather noisy. There are lots of artifacts in the catalogs (e.g. random ink dots) that can be interpreted as the dot symbol. Moreover, sometimes the dot symbol is just missing or almost invisible. Finally, we know the \textit{a priori} position of the dot symbol.

The heatmap is used to decode the sequence of symbols. The most straightforward (but not the only possible as we discuss below) approach is to read the columns of the heatmap from left to right and extract symbols corresponding the highest score in the columns.

Next step is to compress the sequence of symbols into a number. For illustration, consider a shorter sequence: `$\ast\ast--1111\ast776\ast\ast888\ast\ast\ast$'. To obtain a number from this sequence, we first replace the repeated symbols with just one symbol to obtain `$\ast-1\ast76\ast8\ast$' and then eliminate blank symbols to finally obtain the number $-1768$. In most cases we can assume that it is a decimal number with one decimal which results in $-176.8$.

Note that by construction of the CRNN model, the heatmap should not be considered as per column classification of the input image. Moreover, the width of the heatmap is a free parameter and we set it equal to the width of the input image just for convenience. In fact, the only thing that matters for decoding is the order of the columns in the heatmap, not their relative distances.

To train the neural network model, we collect a set of manually labeled images. The dataset contains 8\,000 samples randomly selected from the catalogs. The idea is to provide the model with various examples of handwriting, font colors and background. For training, we use the \textit{connectionist temporal classification} loss function (CTC loss, see \citet{CTC} for more details). We also apply various image augmentation procedures (e.g. randomly change contrast and distort the image) to obtain better generalization capabilities of the model. 

For the trained model we obtain a 97\% accuracy (i.e.\ the percentage of correctly recognized numbers) on the unseen examples of the labeled dataset.  

\begin{figure}
\centering
\includegraphics[width=0.65\textwidth]{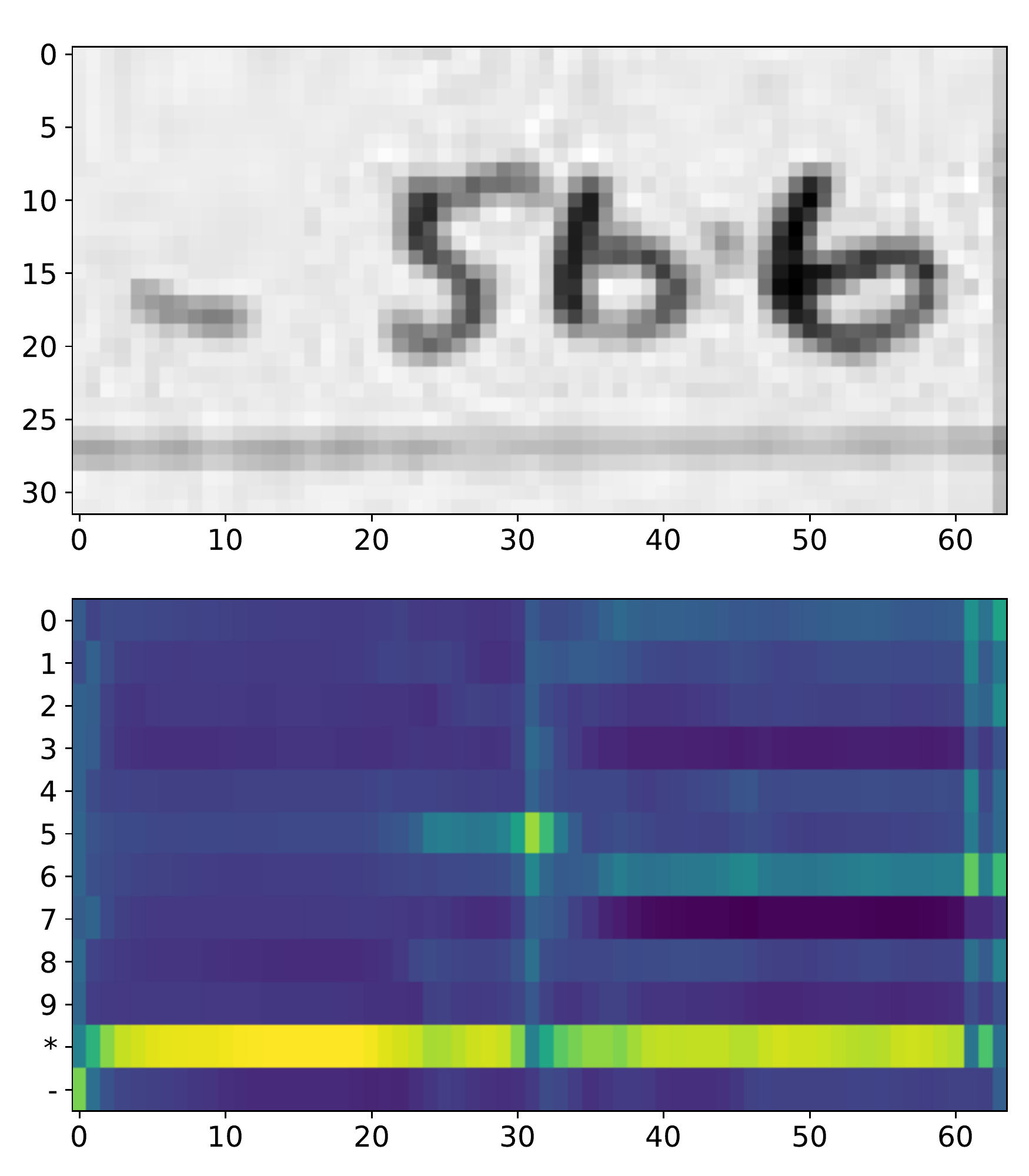}
\caption{Top panel: sample image from the scanned observing book. Bottom panel: heatmap produced by the trained model. Note that the heatmap should not be considered as per column classification of the input image. Only order of the columns in the heatmap is essential for decoding, not their relative distances. Possible decodings of this heatmap are `$-566$', `$-560$', `$566$', `$-562$', `$-506$' with confidences 0.89, 0.08, 0.01, 0.01, 0.01.}
\label{fig:heatmap}
\end{figure}

\begin{figure}
\centering
\includegraphics[width=0.65\textwidth]{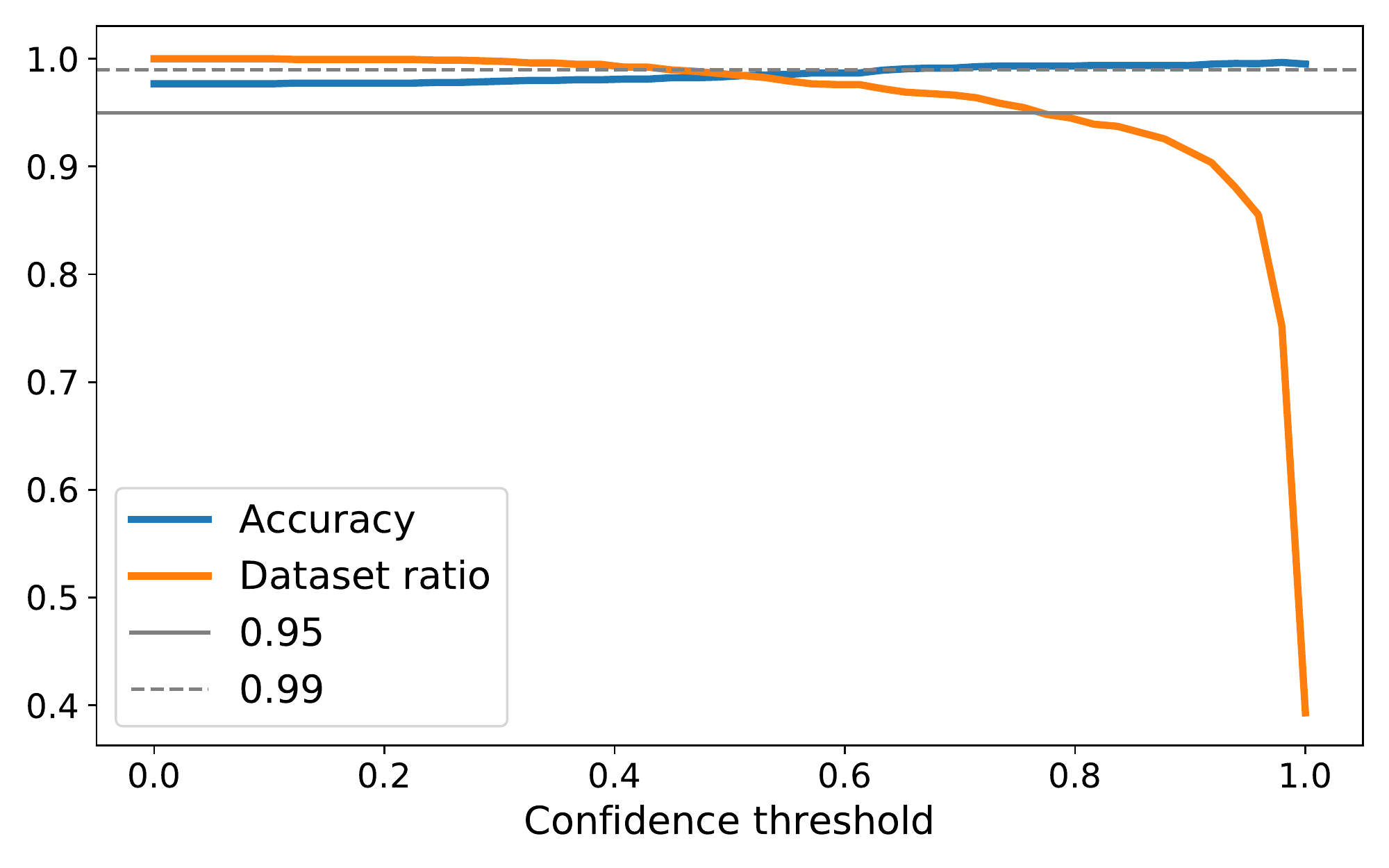}
\caption{The accuracy of the model against the confidence threshold (blue line) and the ratio of samples in the dataset with confidence above the threshold (orange line).}
\label{fig:confidence}
\end{figure}

Note that the heatmap allows confidence estimation and can provide us with multiple possible interpretations. Indeed, instead of selecting a symbol with the highest score, we can consider each column as probability distribution and sample a symbol from this distribution. In fact, we sample 100 sequences from each heatmap and select the five most frequently decoded numbers. The frequency of each number gives us the confidence estimate. Figure~\ref{fig:heatmap} shows an example of the heatmap and ranked interpretations with the confidence estimates. Figure~\ref{fig:confidence} shows how the accuracy of the model depends on the confidence threshold and how many samples in the test set have confidences above this threshold. 

While the model demonstrates quite optimistic performance on the test set, which is 1\,600 samples, this test set is less than 0.05\% of the whole set of unlabeled data, which is about 3.3M. Thus we expect that the model performance can be lower on certain pages with complex conditions (a lot of noise, unexpected background etc) or on previously unseen hand-writings. We assume that deeper models (with larger number of trainable parameters) could provide better generalization, however, they would require substantially larger dataset of labeled examples, which is impractical. Instead, to adjust the model to local conditions, we introduce a \textit{per-page fine-tuning} method described below. However, before applying this method, some auxiliary steps are required and we now proceed to these steps.

\subsection{Font Color Identification\label{color_identification}}

We need to attribute the color of each table row to one of three main colors: blue, red and black. However, we do not know \textit{a priori} where to set the optimal thresholds separating the colors. Moreover, due to the natural ink variation, these thresholds might not be constant in time. We apply clustering in RGB space to define the optimal thresholds for each catalog.

\begin{figure}
\centering
\includegraphics[width=0.85\textwidth]{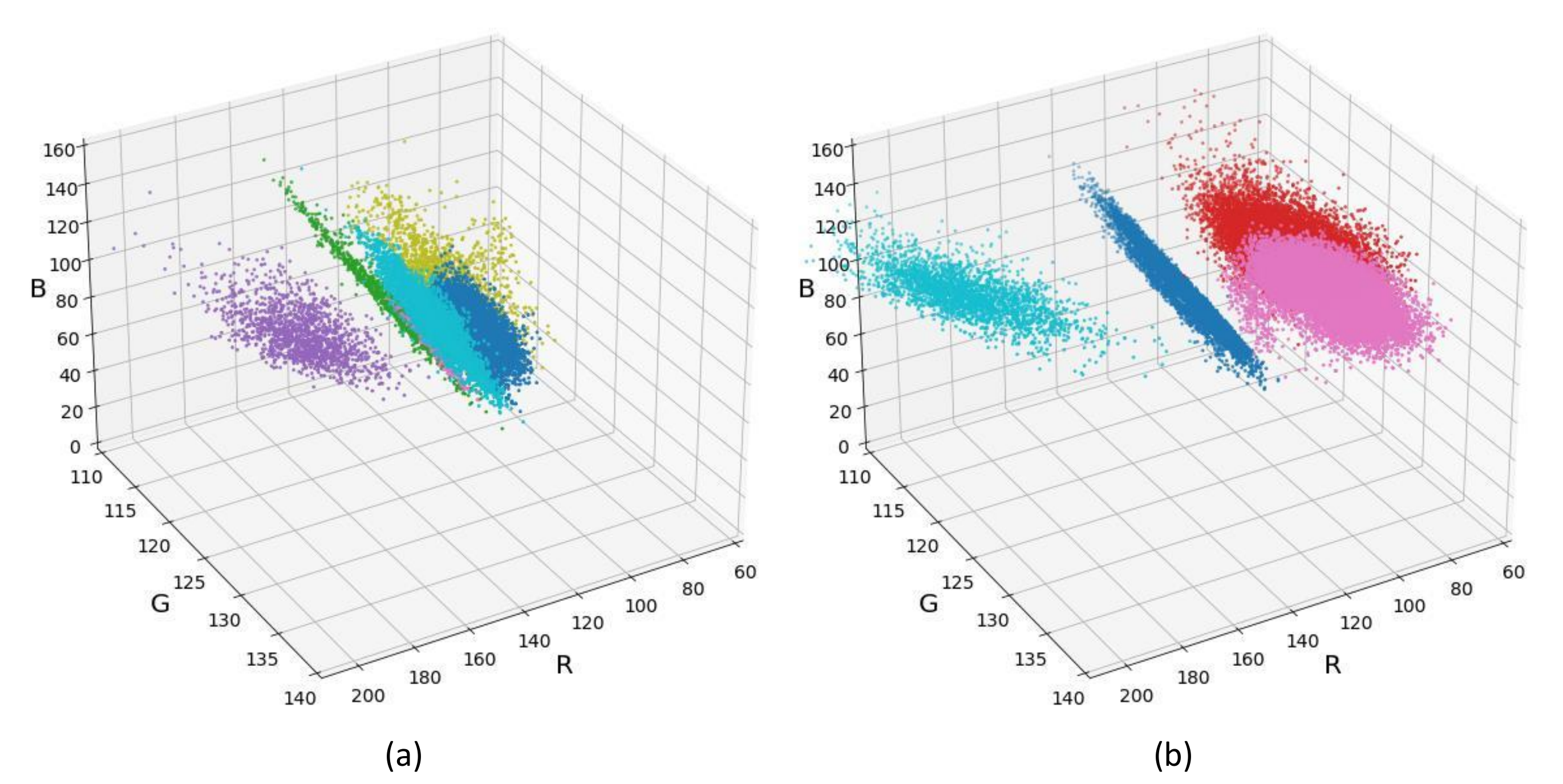}
\caption{Font color distribution in the RGB space obtained for the book 1911--1915 (left panel) and for the book 1916-1920 (right panel). Each dot represents a table record. The colors discriminate clusters isolated with the Gaussian Mixture Model. }
\label{fig:colors}
\end{figure}

More detailed, we first threshold pixel intensities to separate strokes and background. Then we average pixel intensities in a row to obtain a single RGB code for the row. Thus, each row of the table is represented by one point in 3D RGB space. The idea is that we expect to find three clusters corresponding to black, blue and red colors. However, we find it more useful to isolate more clusters and then interpret each as black, blue and red. Figure~\ref{fig:colors} shows color clusters isolated with the Gaussian Mixture Model in the catalogs 1911--1916 and 1916--1920. Note that the position of the clusters varies between the catalogs. Attributing each cluster to one of three main colors, we thereby attribute each row in the catalog to one of three main colors. When we visually verified the color of each row, we made only a few corrections and thus concluded that the method is robust. We repeat this procedure for each catalog.

\subsection{Derivation of the Linear Relationships}

The initial number recognition and color detection allow us to separate series of observations. Now within each series we can extract the most frequent sum of columns $\lambda'$ and $\lambda+k$ (modulo 360) as well as the difference of columns $L$ and $l$ (also modulo 360), since we expect those to be constant for any given moment of time (cf.\ Sect.~\ref{dataset}). Taking into account that the initial number recognition model is good on average, we expect that for most series we derive the linear relationships correctly. At the next step we will use these relationships to improve the recognition of each page.

\subsection{Fine-tuning}

\begin{figure}
\centering
\includegraphics[width=0.75\textwidth]{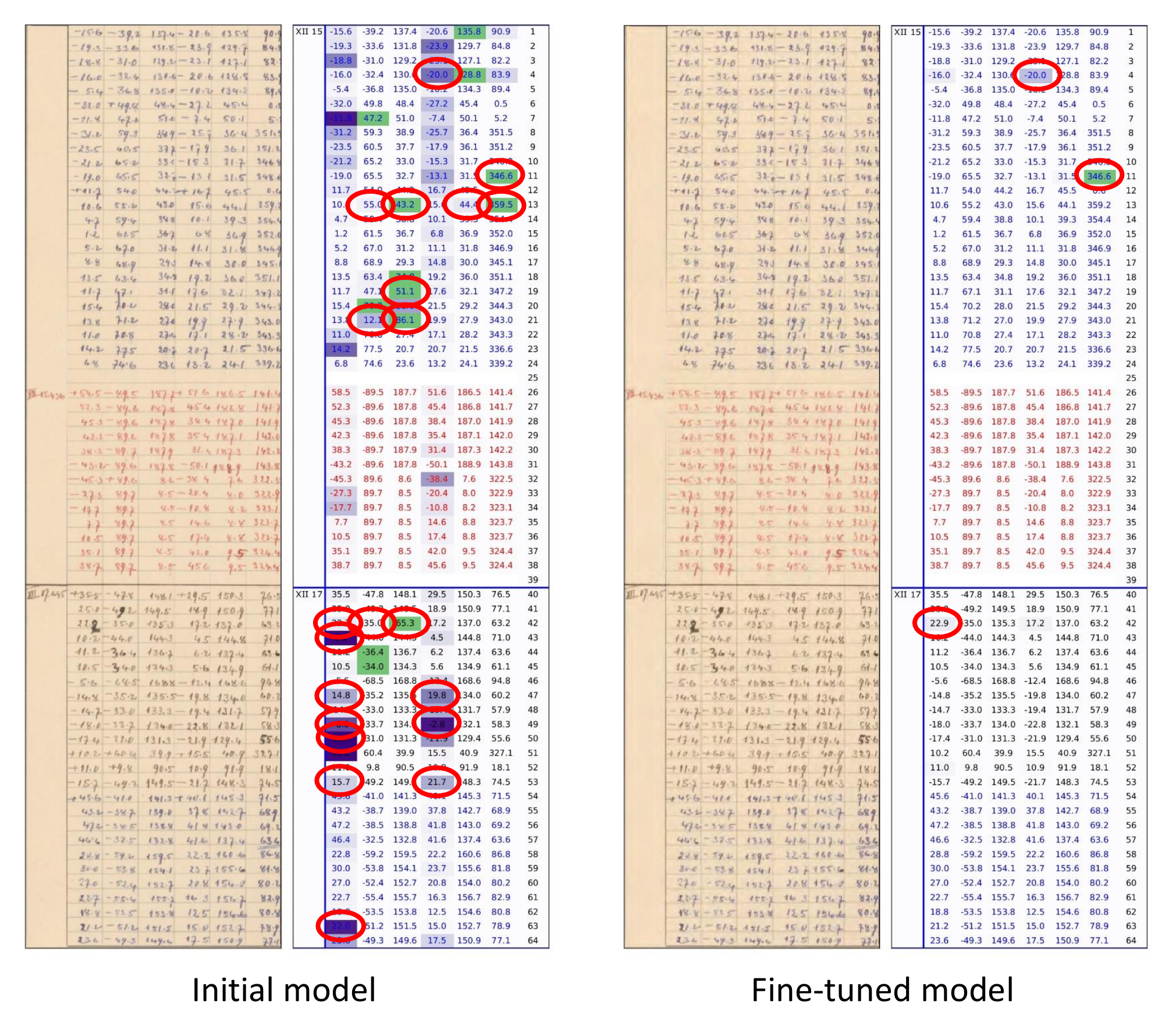}
\caption{Reconstruction of page 1330 from the catalog 1893--1898. Colored cells denote confidences lower than unity (in particular, green background indicates cells where the values were reconstructed using the linear relationships with other columns). Red circles indicate wrong interpretations found by visual verification. \textit{Left panel:} the initial model makes 21 errors (94\% accuracy for this page) and often gives low confidence for correct answers. \textit{Right panel:} the fine-tuning procedure improves the results substantially. The model makes now just 3 errors (99\% accuracy) and confidence in correct answers is close to 1. Note that cells with wrong answers have lower confidence (darker background) as desired. By setting the minimal confidence threshold one can eliminate such cases from further analysis.}
\label{fig:fine}
\end{figure}

Figure~\ref{fig:fine} (left panel) shows a sample page and numbers recognized by the initial model. Accuracy for this page is 94\% that is somehow lower than expected and, moreover, the model gives low confidence for many correct results. The fine-tuning approach is aimed to improve this situation. The idea is to use the known linear relationships between the columns to derive the target labels and make several training iterations on this page. Indeed, using the model predictions for the column $L$ and the known linear relationship between columns $L$ and $l$, we can reconstruct values in the column $l$. Now we can penalize the model if the reconstruction does not match the actual model prediction for the column $l$. The same works in the opposite direction as well and also for the columns $\lambda'$ and $\lambda+k$. We train the model on the page until the reconstruction of one cell from another with a linear relation does not deliver a gain in confidence of the actual output anymore. Figure~\ref{fig:fine} (right panels) shows the results after the fine-tuning step. Now the accuracy is 99\% and almost all correct numbers have confidence close to 1. We repeat this procedure for each page in the catalogs.

\subsection{Post-processing}

Recall that for each cell of the table we obtain a set of decodings with corresponding confidence estimates (more precisely, we consider 5 values ranked by confidence). The final step is to derive the most suitable value from the set of decoded numbers. Note that this is not necessarily the value with the highest confidence. In fact, we select those values that match a set of natural constraints. For example, knowing the linear relationship between columns $L$ and $l$, we select those values that match this relationship. If we fail to find a proper combination for some row, we select the value with the highest confidence and reconstruct the remaining one following the linear constraint. This logic also works for the columns $\lambda'$ and $\lambda+k$. A similar but a bit more complex rule can be derived to conclude on the values $b$ and $\beta$ given $\lambda+k$. If the value matches all the natural constraints, we set the confidence to unity. Note that it does not \textit{guarantee} a correct output and one should expect a small, but nonzero, probability of wrong interpretations. 

Note that using the heatmap one can also estimate the confidence of each symbol in the number. This can be used not only to isolate potentially misrecognized numbers, but also to indicate the uncertain digits in these numbers. However, we assume that for the potential user of this dataset it is more convenient to have a single confidence value for the whole number.

\section{Results}

\begin{figure}
\centering
\includegraphics[width=0.65\textwidth]{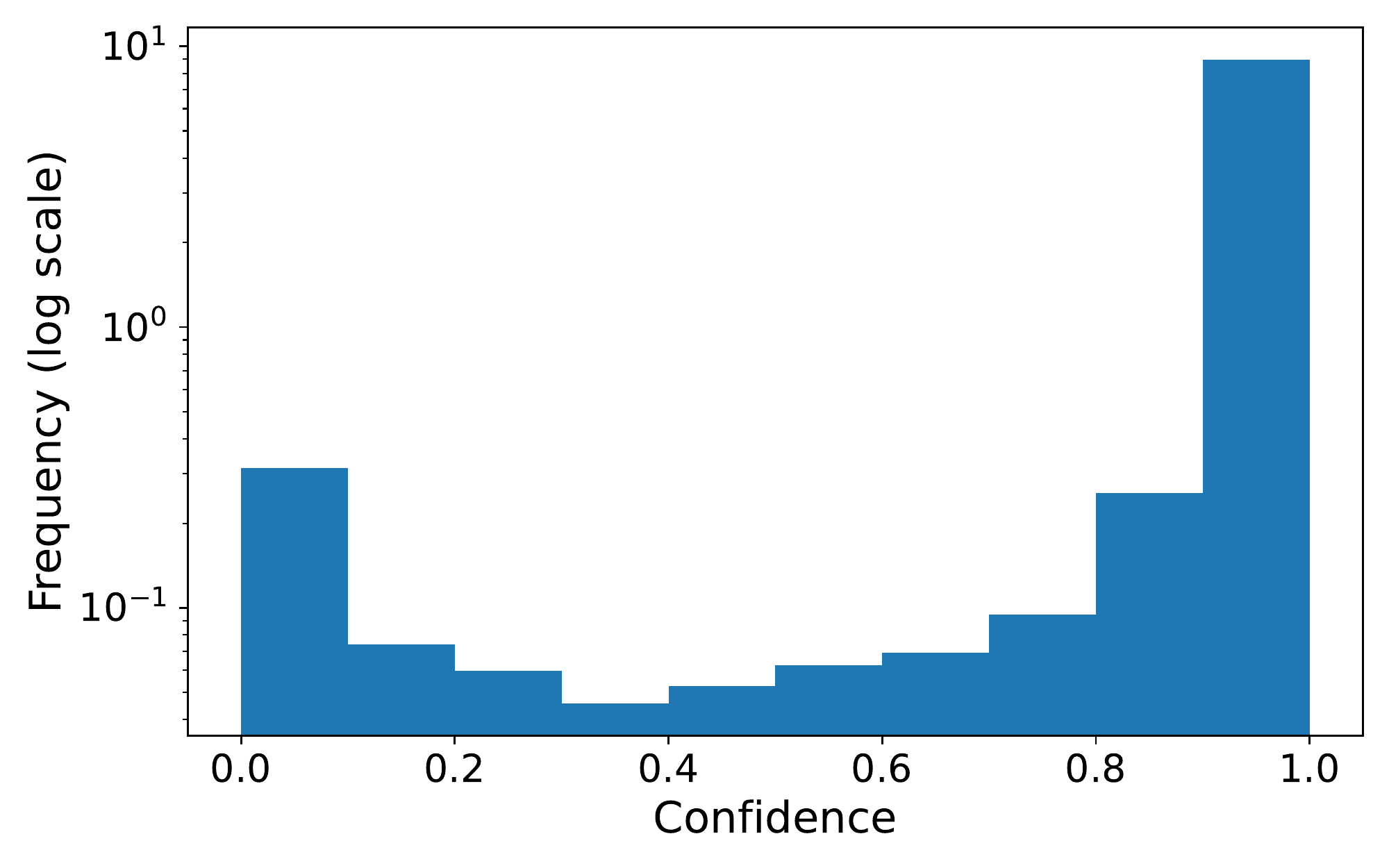}
\caption{Distribution of minimum of confidences for $\beta$, $\lambda'$, $\lambda+k$, $b$, $l$ and $L$ for all records in the catalogs. Note the log scale of the y-axis.}
\label{fig:mconf}
\end{figure}

For further analysis we want to exclude observations which have low confidence (i.e., potentially misrecognized). We define the confidence of the observation as the minimal confidence for values $\beta$, $\lambda'$, $\lambda+k$, $b$, $l$ and $L$. Figure~\ref{fig:mconf} shows the distribution of the minimal confidence. We observe two peaks in the distribution, one near 1 and one near 0 confidence. The value 0.4 looks as an optimal threshold separating the two mixtures. We exclude observations with minimal confidence below 0.4 (this criteria filters out about 5\% of data).

\subsection{Sunspots}

\begin{figure}
\centering
\includegraphics[width=0.85\textwidth]{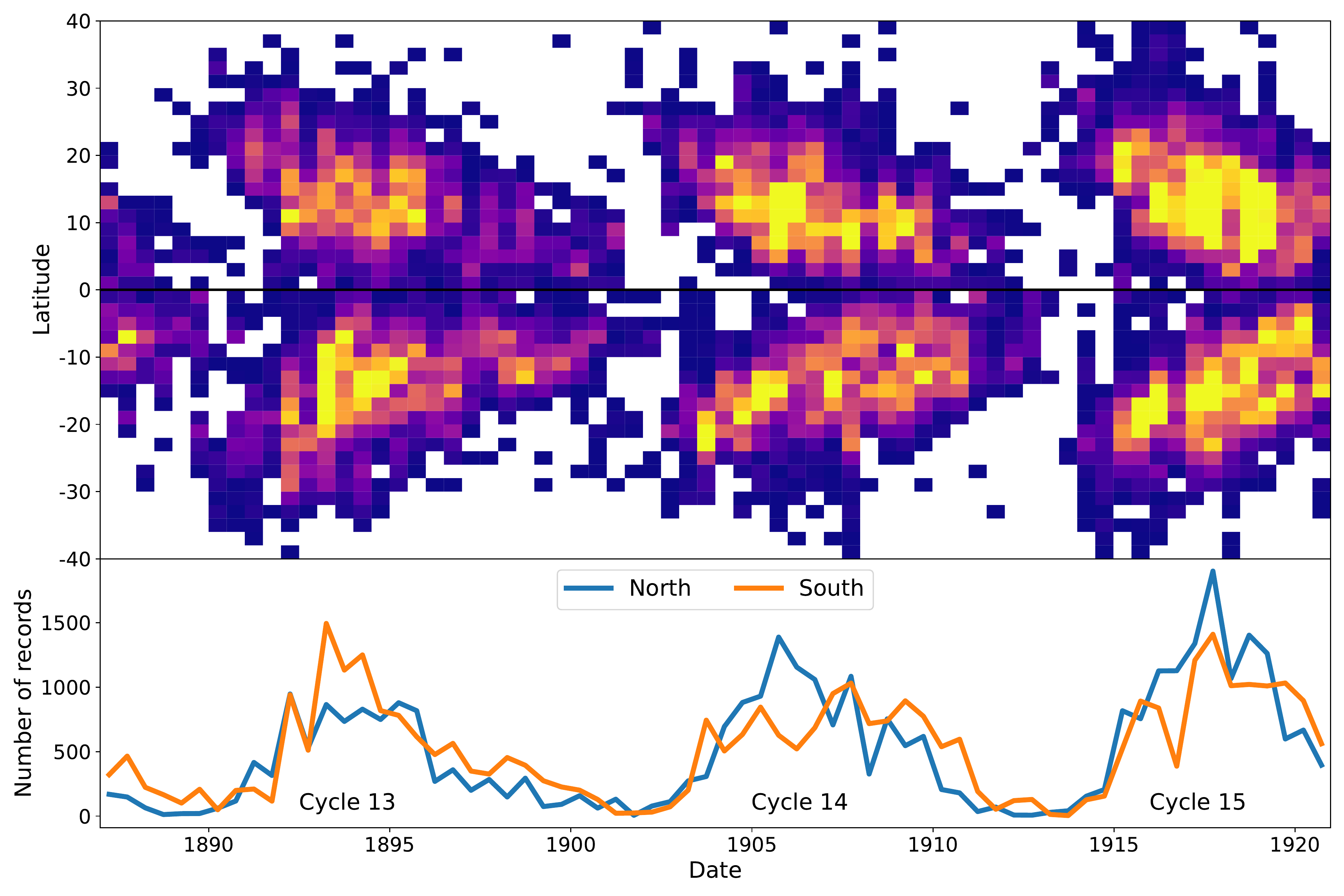}
\caption{\textit{Upper panel:} time-latitude distribution of sunspot records from the Zurich catalogues. Data are binned over 6 months in time and $4^{\circ}$ in latitude. \textit{Bottom panel:} total number of sunspot records in the northern hemisphere (blue line) and in the southern hemisphere (orange line) in 6-month bins.}
\label{fig:sunspots}
\end{figure}

\begin{figure}
\centering
\includegraphics[width=0.85\textwidth]{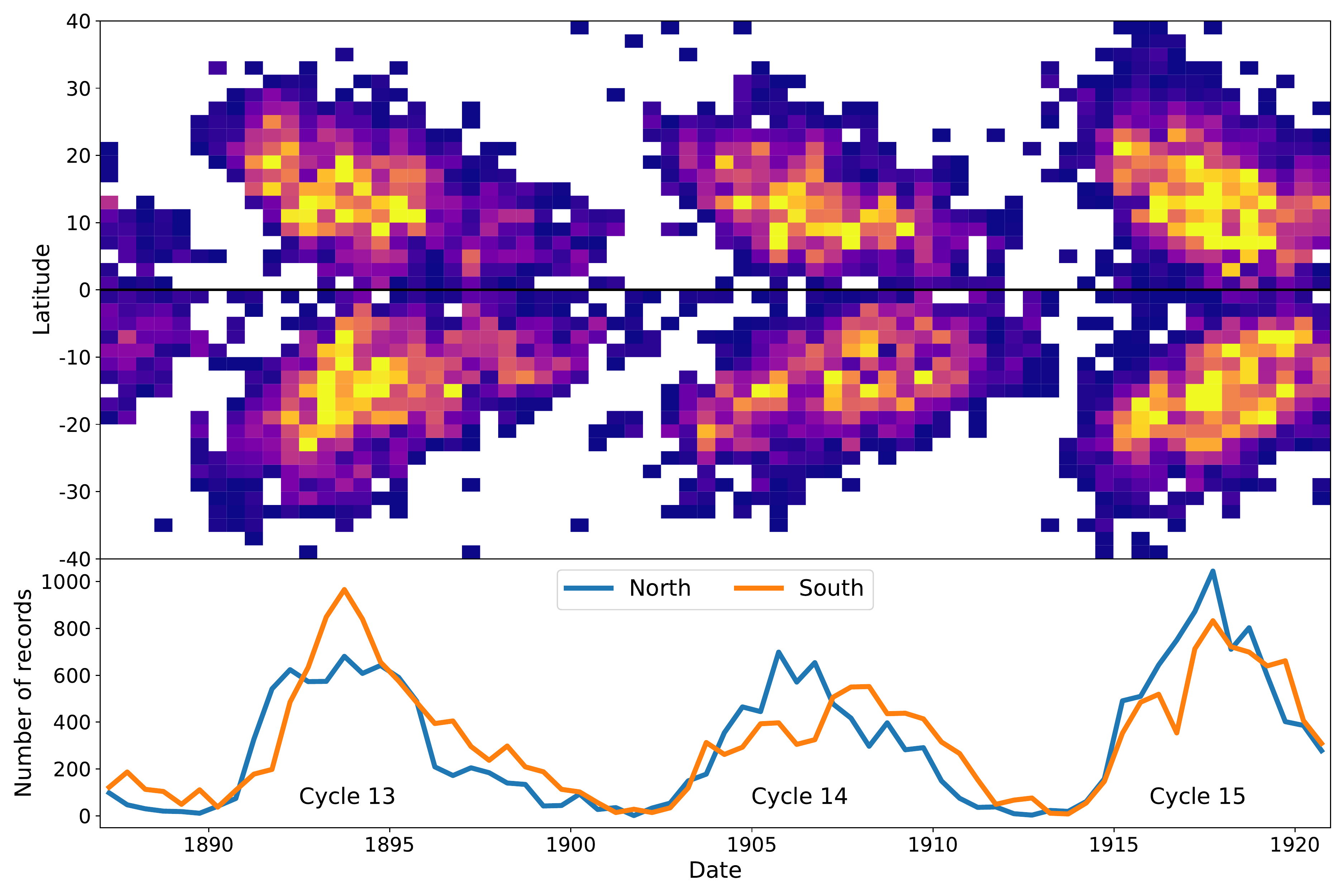}
\caption{Same as Figure~\ref{fig:sunspots} but for sunspot-groups observations from the catalogue of the Royal Greenwich Observatory.}
\label{fig:rgo}
\end{figure}

We identified in total 73\,742~sunspot observations for the period 1887--1920. Filtering out observations with the lowest confidence below 0.4 yields 69\,990~observations. Figure~\ref{fig:sunspots} shows the time-latitude distribution of the database obtained as well as the number of sunspot records in the northern and southern hemispheres in 6-month bins. For comparison, we provide the distribution of sunspot-groups observations according to the catalogue\footnote{\url{https://solarscience.msfc.nasa.gov/greenwch.shtml}} of the Royal Greenwich Observatory (RGO) in Figure~\ref{fig:rgo}.

Both Figures \ref{fig:sunspots} and \ref{fig:rgo} demonstrate a high similarity in the time-latitude distribution, and we conclude that the Zurich catalogs, which provide the positions of individual sunspots, can be used to supplement the widely used catalog of average sunspot-group positions provided by the RGO. There are still some local differences visible in Figures~\ref{fig:sunspots} and~\ref{fig:rgo}. In general, manual verification is required to conclude on particular cases, and we expect that many of these cases will be investigated in a future database study process. We also note that a number of these cases can be eliminated by stronger or more complex sunspot selection criteria (in the simplest case by increasing the minimum confidence threshold).

Comparing the Zurich and RGO catalogs one should take into account that longitudes are not equivalent in the two datasets since different solar rotation speeds are used for the coordinates determination. In order to find a relation between the two catalogs we selected several days in the period 1887--1920 and manually estimated the longitudinal shift between the corresponding observations. We conclude that (i) the reference time and rotation period did not change during these years; (ii) relative speed is estimated as 0.0821~degrees/day and the Zurich has a shorter rotation period; (iii) on 1880 Jan~01 the heliographic longitude of $0^{\circ}$ in the Zurich catalog corresponds to $-68.7055^{\circ}$ in the RGO catalog. Using these parameters we translated coordinates of the Zurich catalogs to match coordinates in the RGO catalogs.

\begin{figure}
\centering
\includegraphics[width=0.95\textwidth]{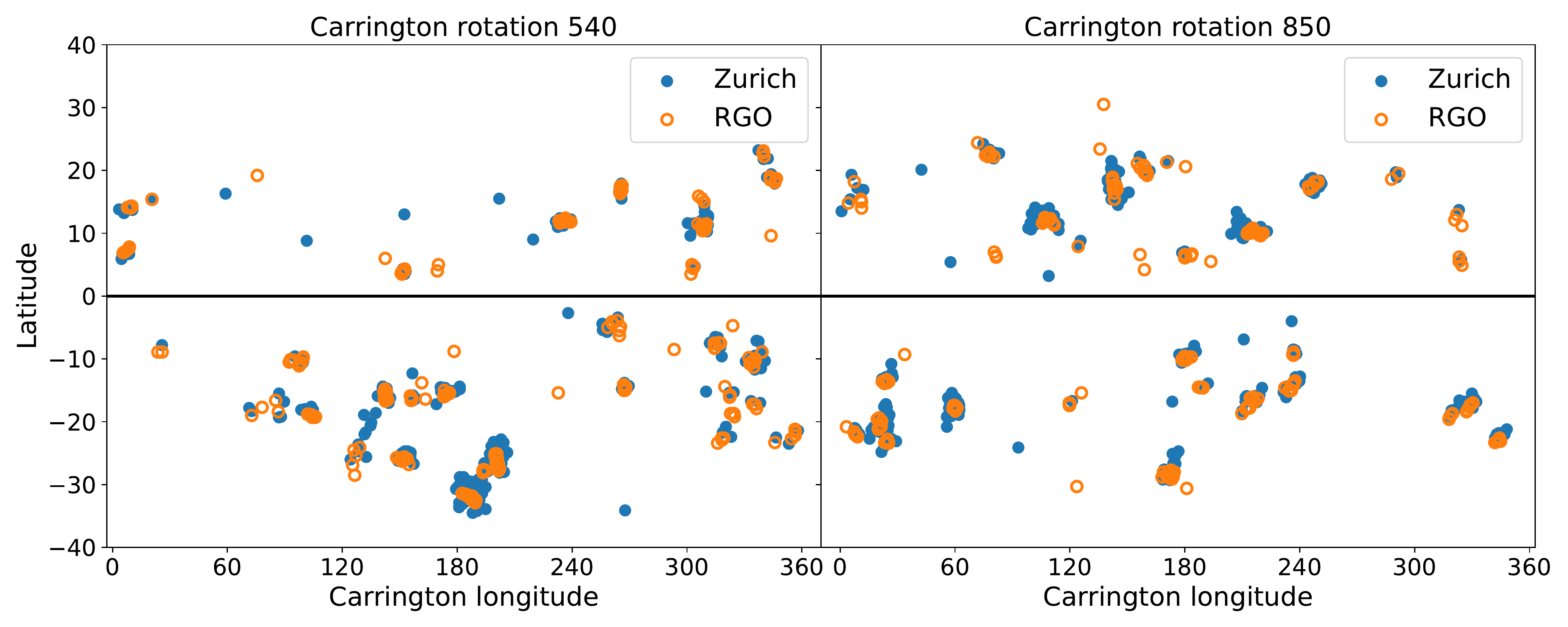}
\caption{Sample synoptic maps with sunspot positions (the blue dots) according to the Zurich catalog and sunspot-group positions (the orange circles) according to the RGO catalog.}
\label{fig:carr}
\end{figure}

In Figure~\ref{fig:carr} we demonstrate the correspondence of the Zurich and RGO catalogs for two sample Carrington rotations, numbers~540 (9~February--8~March 1894) and~850 (5~April--3~May 1917), which are near the maxima of solar Cycles~13 and~15. There is a difference in the completeness of RGO and Zurich sunspot observations. Out of 12\,392~days in the period 1887--1920, RGO data cover 9\,736~days, while Zurich data cover only 6\,710~days. The main reason is that RGO benefits from several observational locations around the world and is therefore less affected by, e.g., weather conditions.

\begin{figure}
\centering
\includegraphics[width=0.65\textwidth]{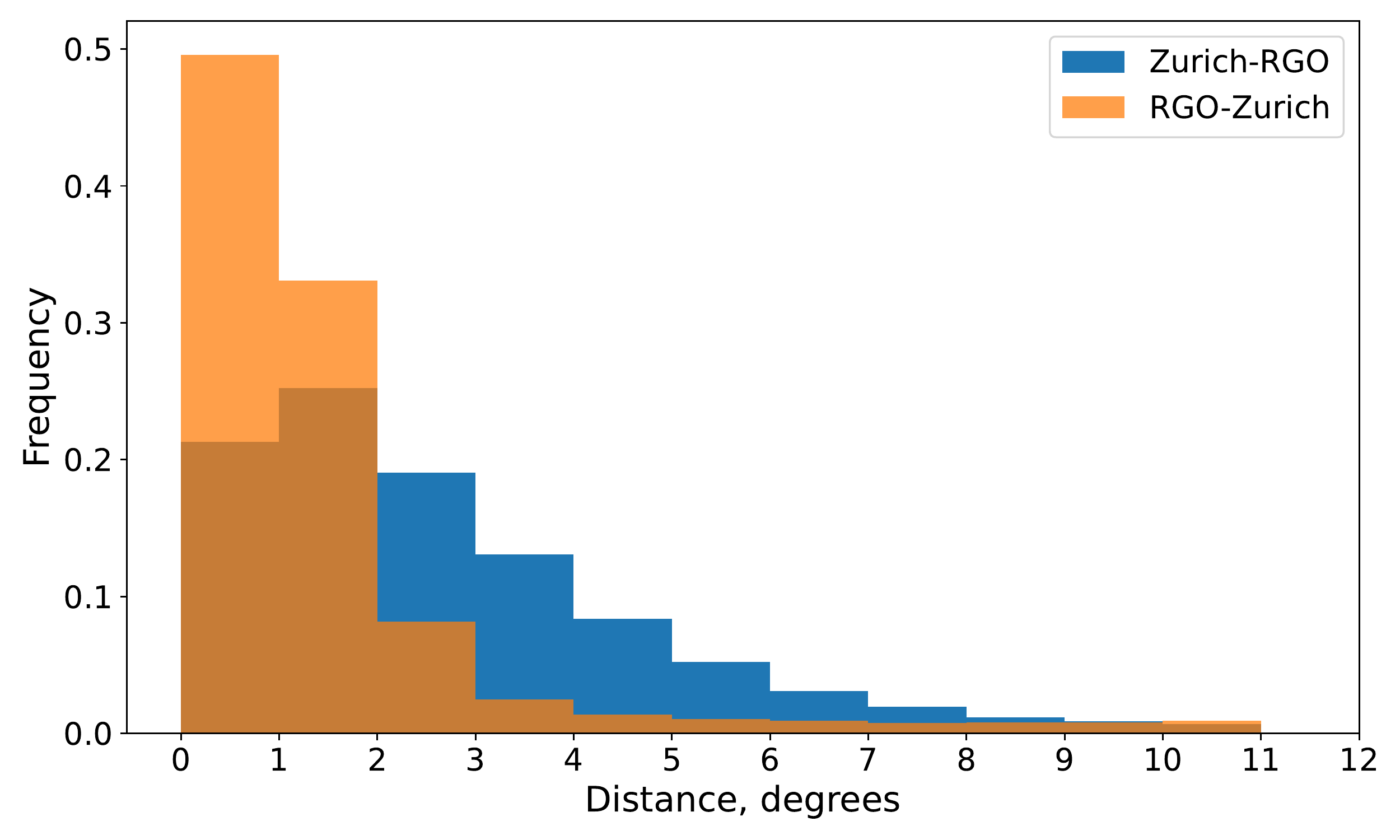}
\caption{Distribution of distances from Zurich sunspots to the nearest sunspot group in the RGO catalog (blue bars) and the distribution of distances from sunspot groups in the RGO catalog to the nearest sunspot in Zurich catalog (orange bars).}
\label{fig:distr}
\end{figure}

In Figure~\ref{fig:distr} we show the distribution of distances from Zurich sunspots to the nearest sunspot group in the RGO catalog and the distribution of distances from sunspot groups in the RGO catalog to the nearest Zurich sunspot. Note that we consider only those days, that are present in both the RGO and Zurich databases. We find that for 93\% of the Zurich sunspots, there is at least one RGO group in a radius of $10^\circ$ \citep[this radius is about half of longitudinal size of the largest groups, see, e.g.,][]{Illarionov2022}. Correspondingly, for 87\% of RGO sunspot groups we find at least one sunspot within $10^\circ$ radius in the Zurich catalogs. Note also that the Zurich-RGO distribution in Figure~\ref{fig:distr} is shifted towards larger distances in comparison to the RGO-Zurich distribution. This feature is rather expected because distances of sunspots in Zurich from the nearest sunspot-group center in RGO will in general be larger than the distance of a given sunspot-group center in RGO from the nearest sunspot in Zurich.

\subsection{Faculae and Prominences}

\begin{figure}
\centering
\includegraphics[width=0.85\textwidth]{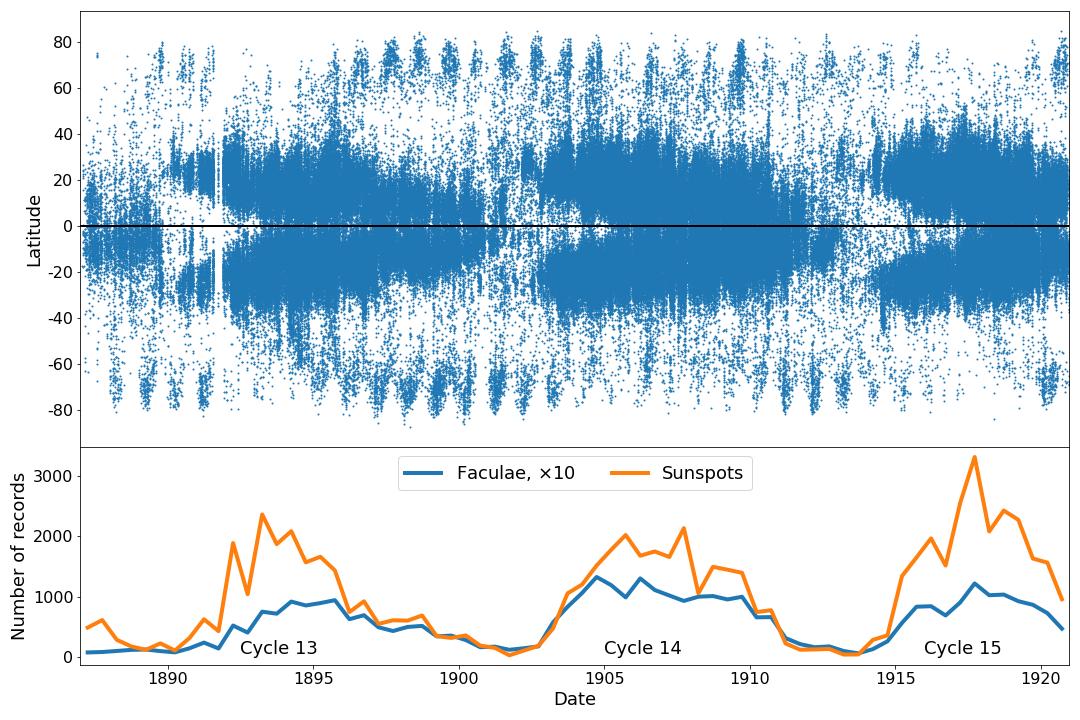}
\caption{\textit{Upper panel:} time-latitude distribution of faculae. \textit{Bottom panel:} total number of faculae (blue line) records in comparison with the total number of sunspot records in the database (orange line).}
\label{fig:facu}
\end{figure}

\begin{figure}
\centering
\includegraphics[width=0.85\textwidth]{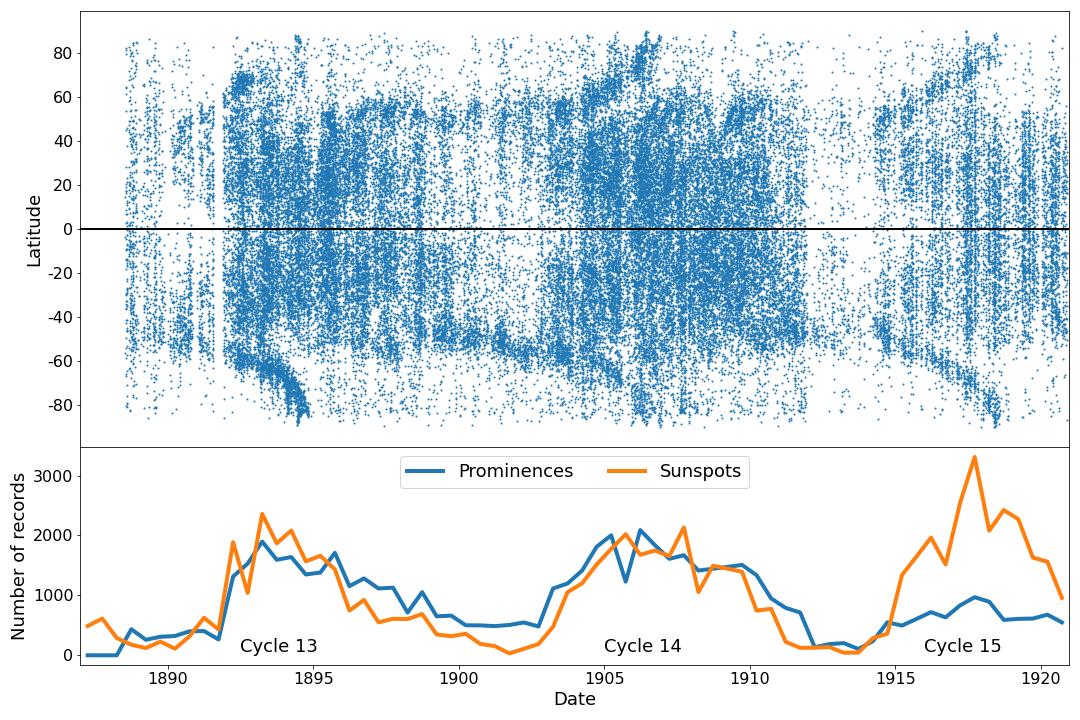}
\caption{Same as Figure~\ref{fig:facu} but for prominences.}
\label{fig:prom}
\end{figure}

We identified in total 413\,098 faculae and 67\,448 prominences for the period 1887--1920. Filtering out observations with the lowest confidence below 0.4 yields 394\,775 faculae and 62\,203 prominences. Figures~\ref{fig:facu} and~\ref{fig:prom} show the corresponding time-latitude distributions and comparisons with the total numbers of faculae and prominences, respectively, as well as of sunspot records. The number of faculae at high latitudes varies in an annually repeated pattern which is caused by the varying visibility of the solar poles. Note how the southern abundance peaks of faculae are shifted by about half a year against the peaks near the north pole. 

\begin{figure}
\centering
\includegraphics[width=0.75\textwidth]{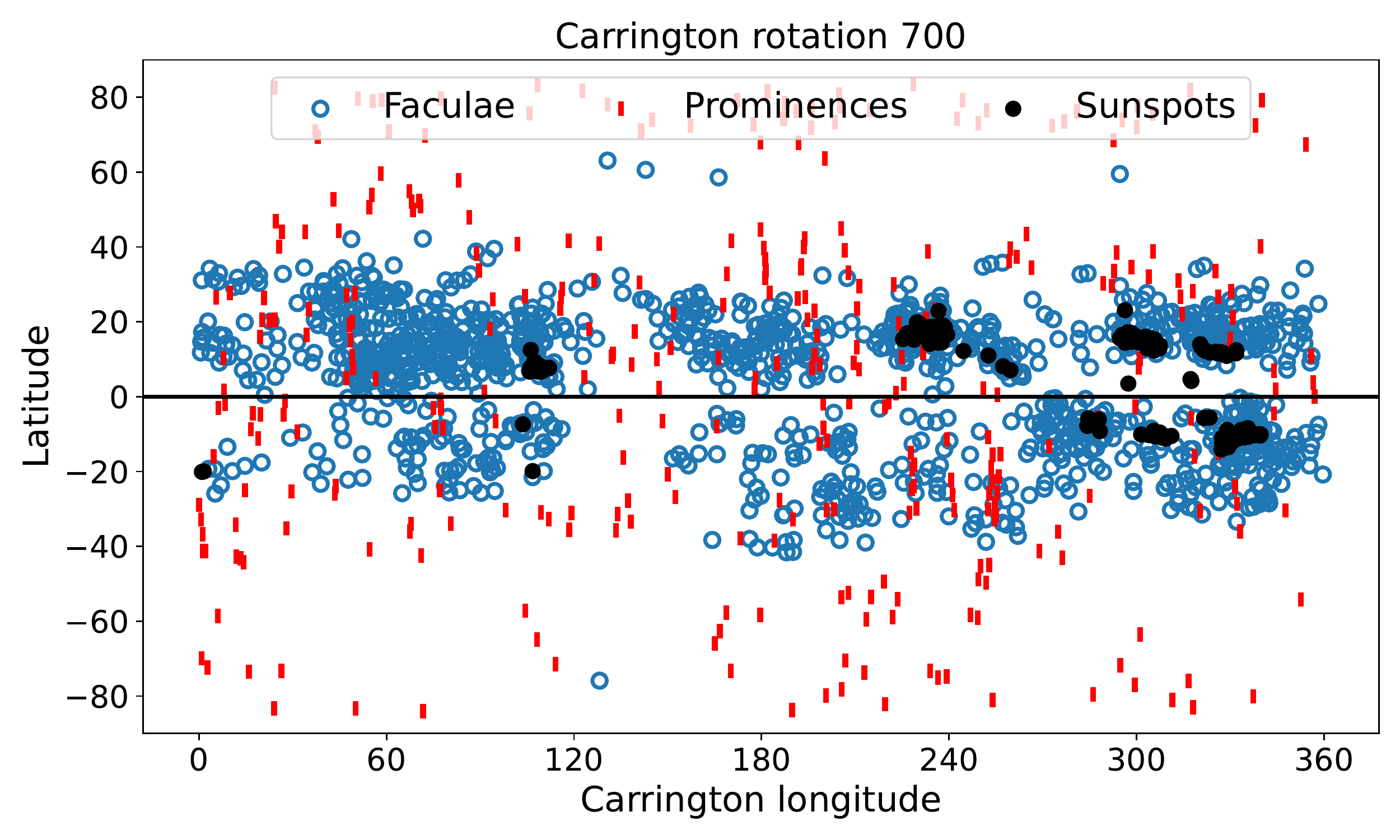}
\caption{Sample synoptic map with sunspots (black dots), prominences (red strips) and faculae (blue circles) overlaid.}
\label{fig:map}
\end{figure}

Note that recently \citet{Tlatova} presented time-latitude diagrams of prominences digitized from Wolfer's synoptic maps for the period 1887--1898 and from the the
solar disk observations in the CaII K-line at the Kodaikanal Solar Observatory for the period  1910--1954. Our work completes these results for the period 1899--1909. Similarly to \citet{Tlatova}, we also observe prominent polar drifts at high latitudes and that the number of prominences correlates with the number of sunspots. Rather surprisingly, the relative number of prominences and sunspots in the Cycle~15 is significantly lower than in the Cycles~13 and~14, for which we observe almost equal numbers of prominences and sunspots. Finally, Figure~\ref{fig:map} shows a sample synoptic map for Carrington rotation 700 (22~January--18~February 1906) taken near the maximum of Cycle~14 with sunspots, prominences and faculae combined.

\section{Discussion and Conclusions}

We presented a neural-network model for the recognition of handwritten coordinates of sunspots, prominences and faculae in catalogs of the Zurich Observatory for the period 1887--1920. The problem turned out to be quite challenging and required the elaboration of a unique approach. Usually, a neural network model is first trained on a large number of labeled examples and then applied with frozen parameters to new data. At first we followed this strategy and obtained an accuracy above 97\% on the test set of labeled data. When we applied the model to the whole set of unlabeled data, we observed that there were certain handwriting and page conditions where the model made many errors and gave low confidence even for correct answers (see Figure~\ref{fig:fine} for the example). This is rather expected since due to the limited size of the training dataset we have to use models with a moderate number of neurons, which in turn have a limited generalization capability. We applied various data augmentations to synthetically extend the dataset, but they can not totally substitute the need for a larger dataset. Of course, it would be interesting for machine learning to develop a universal model equally suitable for any catalog, page condition, and handwriting, probably, at the cost of significant extension of the manually labeled dataset, but it is out of the scope of this research.

Instead, we developed the page-adaptive training approach. We used the fact that for each particular date and time of observation there is a certain relationship between the coordinates presented in the table. The baseline model trained on a set of labeled samples was good enough to recognize a sufficient number of coordinates that allow a reconstruction of these relationships on each new page. Then we used the predictions in one column to compute targets for the second column and \textit{vise versa}. The obtained prediction-target pairs we used to fine-tune the model (i.e. we trained the model further on these pairs). After a certain number of iterations, the model provided substantially better recognition of the whole page in contrast to the baseline model. Repeating this procedure for each new page, we obtained the adaptive model that quickly learned new page conditions and handwritings.

In addition to the number recognition model, we proposed an approach to confidence estimation and generation of a ranked sequence of possible interpretations of each decimal number. Using this approach we identified cases with low confidence and used additional heuristics to decide between several possible interpretations. In general, low confidence can be used as an indicator of possibly incorrect results which require visual verification.

Of course, the algorithm can not guarantee the absence of errors even for results with unity confidence. However, investigating particular cases we find that either the row contains another coordinate with low confidence or there is some systematics in such errors. Both suggest that at least some erroneous records can be eliminated (or even corrected) by implementing some additional heuristics. In this work, we excluded observations with confidences below 0.4 from the analysis of physical properties.

To evaluate the results of coordinates reconstruction we provided the time-latitude diagrams for sunspots, prominences and faculae. In particular, we observed a similar pattern compared with the catalog of sunspot groups provided by the Royal Greenwich Observatory. For a more detailed correlation with the RGO catalog we associate the longitudes in that catalog with the longitudes computed by the Zurich Observatory and obtained a distribution of heliocentric distances between the records in the two catalogs. The scatter in locations is compatible with the scatter of individual sunspots (Zurich) around the center of a sunspot group (RGO).
The observed similarity between the both catalogs suggests both the stability of the  recognition method and the coincidence in the criteria of the observers of both observatories when determining the solar structures. The latter confirms the idea that the Zurich catalogs can complement the RGO data.

We believe that the new database constructed from the observations of the Zurich Observatory will contribute to the still fragmentary knowledge on solar activity in the late 19th and early 20th centuries. We publish the database in the public GitHub repository \url{https://github.com/observethesun/zurich_catalogs} as open-source. The repository is open for everyone to contribute corrections and to report issues. For visual estimation of the recognition accuracy, we provide  catalogs with page-by-page comparisons of the original images and the numerical tables (see the GitHub repository for the actual link). When using the heliographic coordinates, one should be aware of the facts that (i) the longitudes are in a frame with a sidereal rotation period of $25.2339$~days, which is different from the Carrington rotation period of $25.38$~days, and that (ii) the inclination of the solar axis was assumed to be $7^\circ$ upon transforming coordinates by the Zurich assistants, while it should be more like $7.25^\circ$. 
   
\begin{acks}
We greatly acknowledge the efforts by \href{https://e-manuscripta.ch}{e-manuscripta.ch}, the digital platform for manuscript material from Swiss libraries and archives. 
We are also grateful to the group of students of MSU who helped with initial dataset annotation and model validation and the referee for valuable comments. EI acknowledges the support of RSF grant 21-72-20067.
\end{acks}

\section*{Data availability}
The datasets generated during the current study are available in the GitHub repository \url{https://github.com/observethesun/zurich_catalogs}.


  
\bibliographystyle{spr-mp-sola}
\bibliography{literature}

\begin{thebibliography}{16}
\ifx\bisbn     \undefined \def\bisbn  #1{ISBN #1}\fi
\ifx\binits    \undefined \def\binits#1{#1}\fi
\ifx\bauthor   \undefined \def\bauthor#1{#1}\fi
\ifx\batitle   \undefined \def\batitle#1{#1}\fi
\ifx\bjtitle   \undefined \def\bjtitle#1{\textit{#1}}\fi
\ifx\bvolume   \undefined \def\bvolume#1{\textbf{#1}}\fi
\ifx\byear     \undefined \def\byear#1{#1}\fi
\ifx\bissue    \undefined \def\bissue#1{#1}\fi
\ifx\bfpage    \undefined \def\bfpage#1{#1}\fi
\ifx\blpage    \undefined \def\blpage #1{#1}\fi
\ifx\burl      \undefined \def\burl#1{#1}\fi
\ifx\href      \undefined \def\href#1#2{#2}\fi
\ifx\betal     \undefined \def\betal{et al.}\fi
\ifx\bctitle   \undefined \def\bctitle#1{#1}\fi
\ifx\beditor   \undefined \def\beditor#1{#1}\fi
\ifx\bbtitle   \undefined \def\bbtitle#1{\textit{#1}}\fi
\ifx\bedition  \undefined \def\bedition#1{#1}\fi
\ifx\bseriesno \undefined \def\bseriesno#1{\textbf{#1}}\fi
\ifx\blocation \undefined \def\blocation#1{#1}\fi
\ifx\bsertitle \undefined \def\bsertitle#1{\textit{#1}}\fi
\ifx\bsnm      \undefined \def\bsnm#1{#1}\fi
\ifx\bsuffix   \undefined \def\bsuffix#1{#1}\fi
\ifx\bparticle \undefined \def\bparticle#1{#1}\fi
\ifx\barticle  \undefined \def\barticle#1{}\fi
\ifx\binstitute  \undefined \def\binstitute#1{#1}\fi
\ifx\bpublisher  \undefined \def\bpublisher#1{#1}\fi
\ifx\doiurl    \undefined \def\doiurl#1{\href{#1}{DOI}}\fi
\makeatletter
\def\safeHref#1#2#3{\in@{http}{#2}\ifin@\href{#2}{#3}\else\href{#1#2}{#3}\fi}
\makeatother
\ifx\adsurl    \undefined
  \def\adsurl#1{\safeHref{https://ui.adsabs.harvard.edu/abs/}{#1}{ADS}}\fi
\ifx\arxivurl  \undefined
  \def\arxivurl#1{\safeHref{http://arxiv.org/abs/}{#1}{arXiv}}\fi
\ifx\botherref \undefined \def\botherref#1{}\fi
\ifx\url       \undefined \def\url#1{#1}\fi
\ifx\bchapter  \undefined \def\bchapter#1{}\fi
\ifx\bbook     \undefined \def\bbook#1{}\fi
\ifx\bcomment  \undefined \def\bcomment#1{#1}\fi
\ifx\oauthor   \undefined \def\oauthor#1{#1}\fi
\ifx\citeauthoryear \undefined\def \citeauthoryear#1{#1}\fi
\def\endbibitem {}
\ifx\bconflocation  \undefined \def\bconflocation#1{#1} \fi

\bibitem[\protect\citeauthoryear{{Arlt} and {Vaquero}}{2020}]{arlt_vaquero2020}
\begin{barticle}
\bauthor{\bsnm{{Arlt}}, \binits{R.}},
\bauthor{\bsnm{{Vaquero}}, \binits{J.M.}}:
\byear{2020},
\batitle{{Historical sunspot records}}.
\bjtitle{Living Reviews in Solar Physics}
\bvolume{17},
\bfpage{1}.
\doiurl{https://doi.org/10.1007/s41116-020-0023-y}.
\adsurl{2020LRSP...17....1A}.
\end{barticle}
\endbibitem

\bibitem[\protect\citeauthoryear{Canny}{1986}]{Canny86}
\begin{barticle}
\bauthor{\bsnm{Canny}, \binits{J.}}:
\byear{1986},
\batitle{A Computational Approach to Edge Detection}.
\bjtitle{IEEE Transactions on Pattern Analysis and Machine Intelligence}
\bvolume{PAMI-8},
\bfpage{679}.
\doiurl{https://doi.org/10.1109/TPAMI.1986.4767851}.
\end{barticle}
\endbibitem

\bibitem[\protect\citeauthoryear{{Clette} et~al.}{2014}]{cliver_ea2014}
\begin{barticle}
\bauthor{\bsnm{{Clette}}, \binits{F.}},
\bauthor{\bsnm{{Svalgaard}}, \binits{L.}},
\bauthor{\bsnm{{Vaquero}}, \binits{J.M.}},
\bauthor{\bsnm{{Cliver}}, \binits{E.W.}}:
\byear{2014},
\batitle{{Revisiting the Sunspot Number. A 400-Year Perspective on the Solar
  Cycle}}.
\bjtitle{\ssr}
\bvolume{186},
\bfpage{35}.
\doiurl{https://doi.org/10.1007/s11214-014-0074-2}.
\adsurl{2014SSRv..186...35C}.
\end{barticle}
\endbibitem

\bibitem[\protect\citeauthoryear{{Cliver}}{2017}]{cliver2017}
\begin{barticle}
\bauthor{\bsnm{{Cliver}}, \binits{E.W.}}:
\byear{2017},
\batitle{{Sunspot number recalibration: The 1840-1920 anomaly in the observer
  normalization factors of the group sunspot number}}.
\bjtitle{Journal of Space Weather and Space Climate}
\bvolume{7},
\bfpage{A12}.
\doiurl{https://doi.org/10.1051/swsc/2017010}.
\adsurl{2017JSWSC...7A..12C}.
\end{barticle}
\endbibitem

\bibitem[\protect\citeauthoryear{{Curto} et~al.}{2016}]{curto_ea2016}
\begin{barticle}
\bauthor{\bsnm{{Curto}}, \binits{J.J.}},
\bauthor{\bsnm{{Sol{\'e}}}, \binits{J.G.}},
\bauthor{\bsnm{{Genesc{\`a}}}, \binits{M.}},
\bauthor{\bsnm{{Blanca}}, \binits{M.J.}},
\bauthor{\bsnm{{Vaquero}}, \binits{J.M.}}:
\byear{2016},
\batitle{{Historical Heliophysical Series of the Ebro Observatory}}.
\bjtitle{\solphys}
\bvolume{291},
\bfpage{2587}.
\doiurl{https://doi.org/10.1007/s11207-016-0896-z}.
\adsurl{2016SoPh..291.2587C}.
\end{barticle}
\endbibitem

\bibitem[\protect\citeauthoryear{{Diercke}, {Arlt}, and
  {Denker}}{2015}]{diercke_ea2015}
\begin{barticle}
\bauthor{\bsnm{{Diercke}}, \binits{A.}},
\bauthor{\bsnm{{Arlt}}, \binits{R.}},
\bauthor{\bsnm{{Denker}}, \binits{C.}}:
\byear{2015},
\batitle{{Digitization of sunspot drawings by Sp{\"o}rer made in 1861-1894}}.
\bjtitle{Astronomische Nachrichten}
\bvolume{336},
\bfpage{53}.
\doiurl{https://doi.org/10.1002/asna.201412138}.
\adsurl{2015AN....336...53D}.
\end{barticle}
\endbibitem

\bibitem[\protect\citeauthoryear{Graves, Fernández, and Gomez}{2006}]{CTC}
\begin{bchapter}
\bauthor{\bsnm{Graves}, \binits{A.}},
\bauthor{\bsnm{Fernández}, \binits{S.}},
\bauthor{\bsnm{Gomez}, \binits{F.}}:
\byear{2006},
\bctitle{Connectionist temporal classification: Labelling unsegmented sequence
  data with recurrent neural networks}.
In: \bbtitle{In Proceedings of the International Conference on Machine
  Learning, ICML 2006},
\bfpage{369}.
\end{bchapter}
\endbibitem

\bibitem[\protect\citeauthoryear{Illarionov and Tlatov}{2022}]{Illarionov2022}
\begin{barticle}
\bauthor{\bsnm{Illarionov}, \binits{E.}},
\bauthor{\bsnm{Tlatov}, \binits{A.}}:
\byear{2022},
\batitle{Parametrization of Sunspot Groups Based on Machine-Learning Approach}.
\bjtitle{Solar Physics}
\bvolume{297}.
\doiurl{https://doi.org/10.1007/s11207-022-01955-0}.
\end{barticle}
\endbibitem

\bibitem[\protect\citeauthoryear{Lepshokov, Tlatov, and
  Vasil'eva}{2012}]{Lepshokov2012}
\begin{barticle}
\bauthor{\bsnm{Lepshokov}, \binits{D.K.}},
\bauthor{\bsnm{Tlatov}, \binits{A.G.}},
\bauthor{\bsnm{Vasil'eva}, \binits{V.V.}}:
\byear{2012},
\batitle{Reconstruction of sunspot characteristics for 1853{\textendash}1879}.
\bjtitle{Geomagnetism and Aeronomy}
\bvolume{52},
\bfpage{843}.
\doiurl{https://doi.org/10.1134/s0016793212070109}.
\end{barticle}
\endbibitem

\bibitem[\protect\citeauthoryear{{Shi}, {Bai}, and {Yao}}{2015}]{crnn}
\begin{botherref}
\oauthor{\bsnm{{Shi}}, \binits{B.}},
\oauthor{\bsnm{{Bai}}, \binits{X.}},
\oauthor{\bsnm{{Yao}}, \binits{C.}}:
2015,
{An End-to-End Trainable Neural Network for Image-based Sequence Recognition
  and Its Application to Scene Text Recognition}.
\textit{arXiv e-prints},
arXiv:1507.05717.
\adsurl{2015arXiv150705717S}.
\end{botherref}
\endbibitem

\bibitem[\protect\citeauthoryear{{Sp\"orer}}{1874}]{spoerer1874}
\begin{bbook}
\bauthor{\bsnm{{Sp\"orer}}, \binits{G.}}:
\byear{1874},
\bbtitle{{Beobachtungen der Sonnenflecken zu Anclam}},
\bpublisher{Leipzig: Wilh. Engelmann}.
\end{bbook}
\endbibitem

\bibitem[\protect\citeauthoryear{{Tlatova}, {Vasil'eva}, and
  {Tlatov}}{2020}]{Tlatova}
\begin{barticle}
\bauthor{\bsnm{{Tlatova}}, \binits{K.A.}},
\bauthor{\bsnm{{Vasil'eva}}, \binits{V.V.}},
\bauthor{\bsnm{{Tlatov}}, \binits{A.G.}}:
\byear{2020},
\batitle{{Drift of Polar Prominences in Solar Cycles 13-24}}.
\bjtitle{Geomagnetism and Aeronomy}
\bvolume{59},
\bfpage{1022}.
\doiurl{https://doi.org/10.1134/S001679321908022X}.
\adsurl{2020Ge&Ae..59.1022T}.
\end{barticle}
\endbibitem

\bibitem[\protect\citeauthoryear{{Watson} and
  {Fletcher}}{2011}]{watson_fletcher2011}
\begin{bchapter}
\bauthor{\bsnm{{Watson}}, \binits{F.}},
\bauthor{\bsnm{{Fletcher}}, \binits{L.}}:
\byear{2011},
\bctitle{{Automated sunspot detection and the evolution of sunspot magnetic
  fields during solar cycle 23}}.
In: \beditor{\bsnm{{Prasad Choudhary}}, \binits{D.}},
\beditor{\bsnm{{Strassmeier}}, \binits{K.G.}} (eds.)
\bbtitle{Physics of Sun and Star Spots}
\bseriesno{273},
\bfpage{51}.
\doiurl{https://doi.org/10.1017/S1743921311014992}.
\adsurl{2011IAUS..273...51W}.
\end{bchapter}
\endbibitem

\bibitem[\protect\citeauthoryear{{Willis} et~al.}{2013}]{willis_ea2013}
\begin{barticle}
\bauthor{\bsnm{{Willis}}, \binits{D.M.}},
\bauthor{\bsnm{{Coffey}}, \binits{H.E.}},
\bauthor{\bsnm{{Henwood}}, \binits{R.}},
\bauthor{\bsnm{{Erwin}}, \binits{E.H.}},
\bauthor{\bsnm{{Hoyt}}, \binits{D.V.}},
\bauthor{\bsnm{{Wild}}, \binits{M.N.}},
\bauthor{\bsnm{{Denig}}, \binits{W.F.}}:
\byear{2013},
\batitle{{The Greenwich Photo-heliographic Results (1874 - 1976): Summary of
  the Observations, Applications, Datasets, Definitions and Errors}}.
\bjtitle{\solphys}
\bvolume{288},
\bfpage{117}.
\doiurl{https://doi.org/10.1007/s11207-013-0311-y}.
\adsurl{2013SoPh..288..117W}.
\end{barticle}
\endbibitem

\bibitem[\protect\citeauthoryear{{Wolf}}{1881}]{wolf1881}
\begin{barticle}
\bauthor{\bsnm{{Wolf}}, \binits{R.}}:
\byear{1881},
\batitle{{Astronomische Mittheilungen LIII}}.
\bjtitle{Astronomische Mitteilungen der Eidgen{\"o}ssischen Sternwarte Zurich}
\bvolume{6},
\bfpage{69}.
\adsurl{1881MiZur...6...69W}.
\end{barticle}
\endbibitem

\bibitem[\protect\citeauthoryear{{Wolf}}{1887}]{wolf1887}
\begin{barticle}
\bauthor{\bsnm{{Wolf}}, \binits{R.}}:
\byear{1887},
\batitle{{Astronomische Mitteilungen LXX}}.
\bjtitle{Astronomische Mitteilungen der Eidgen{\"o}ssischen Sternwarte Zurich}
\bvolume{7},
\bfpage{369}.
\adsurl{1887MiZur...7..369W}.
\end{barticle}
\endbibitem

\end{thebibliography}

\end{article} 
\end{document}